\documentclass{article}
\usepackage{PRIMEarxiv}
\usepackage{amsmath}
\usepackage[utf8]{inputenc} 
\usepackage[T1]{fontenc}    
\usepackage{hyperref}       
\usepackage{url}            
\usepackage{booktabs}       
\usepackage{amsfonts}       
\usepackage{nicefrac}       
\usepackage{microtype}      
\usepackage{lipsum}
\usepackage{fancyhdr}       
\usepackage{cite}
\usepackage{graphicx}       
\graphicspath{{media/}}     

\title{Single-photon and two-photon blockade in a three-wave mixing system with a two-level atom
\thanks{\textit{\underline{Citation}}: 
\textbf{Authors. Title. Pages.... DOI:000000/11111.}} 
}

\author{
  Hong-Yu Lin \\
  College of Physics and electronic information, Baicheng Normal University, Baicheng-137000 \\
  \texttt{lin527668281@126.com} \\
}

\begin{document}
\maketitle

\begin{abstract}
This paper discusses conventional photon blockade (CPB) and two-photon blockade (2PB) in a three-wave mixing system embedded with a two-level atom in the high-frequency cavity. Analytical conditions for achieving CPB and 2PB are obtained by analyzing the eigenvalues of the system Hamiltonian. Numerical solutions, derived by solving the master equation in a truncated Fock space, are consistent with the analytical conditions. Detailed analysis of system parameters reveals the influence of the embedded atom on achieving different types of photon blockade. Unlike previous schemes, this system can achieve single-photon blockade simultaneously in three photon modes. Additionally, by adjusting the coupling coefficient between the atom and high-frequency mode photons, the system can switch between single-photon blockade and two-photon blockade in the high-frequency mode.
\end{abstract}

\keywords{Conventional photon blockade \and Two-photon blockade }

\section{Introduction}
The preparation and manipulation of single photons are currently hot research topics in the fields of quantum optics and quantum information \cite{1,2,3,4,5,6,7,8,9,10,11,12,13}. Consequently, single-photon source devices have become a focal point in these areas. A single-photon source is a typical sub-Poissonian light source, primarily realized by introducing a nonlinear medium into an optical microcavity to achieve the photon blockade (PB) effect \cite{14,15,16,17,18,19,20}. As a non-classical optical effect, PB has garnered widespread attention in theoretical research and experimental implementation due to its potential applications in information technology and communications.
Currently, the realization of PB is mainly focused on two methods: one is conventional photon blockade (CPB), achieved through energy level splitting caused by high nonlinearity in the system; the other is unconventional photon blockade (UPB), primarily caused by quantum interference between different paths. These two methods each demonstrate unique realization mechanisms and application prospects.

In a nonlinear optical system, the strong interaction between photons changes the energy level gradient of the resonator. Although the frequency of the driving field can be tuned to resonate with the cavity, the presence of strong nonlinearity adjusts the distribution of the photon number, causing it to deviate from Poissonian statistics and tend towards sub-Poissonian statistics. The nonlinear coupling in the system alters the original energy level structure, suppressing the probability of multiple photons coexisting. When the nonlinearity is strong enough, the presence of one photon in the cavity prevents the generation of the next photon, which is the physical mechanism of CPB \cite{21,22}. CPB was first observed in an optical cavity coupled with a single trapped atom \cite{23}, and has since been realized in cavity quantum electrodynamics \cite{13,24,26}, optomechanical systems \cite{27,28,29,30}, quantum dots \cite{31}, superconducting circuits \cite{32}, and second-order nonlinear systems \cite{33,34,35}. The realization of UPB mainly occurs because photons can reach a specific photonic state through multiple paths, with quantum interference between different paths achieving this effect \cite{7,17,36,37,38,41,42,43,44}. UPB has been observed in quantum dot systems\cite{45}, superconducting systems \cite{46}, and second-order nonlinear systems. In fact, there is a certain correlation between the physical mechanisms of CPB and UPB \cite{47,48}, and recent studies have shown that CPB and UPB can coexist in some systems \cite{49,50}. Besides being used to prepare single-photon sources, single-photon blockade also has applications in photonic transistors \cite{51}, interferometers \cite{52}, and quantum optical diodes \cite{34}.

With the deepening research on single-photon blockade, the concept of n-photon blockade (nPB) has been proposed. Similar to single-photon blockade, multi-photon blockade typically occurs in nonlinear optical systems, where the generation of n photons inhibits or blocks the subsequent generation and transmission of photons. Currently, the study of two-photon blockade (2PB) is the most extensive, with 2PB being observed in Kerr nonlinear systems \cite{54}, strongly coupled qubit-cavity systems \cite{55,56}, and cascaded cavity quantum electrodynamics systems \cite{57}. Moreover, 2PB can also be achieved through squeezing methods \cite{58}.

This paper studies CPB and 2PB in a three-wave mixing system with a high-frequency cavity embedded with a two-level atom. We analytically analyzed the eigenvalues of the system Hamiltonian and obtained the analytical conditions for realizing CPB and 2PB. We also discussed in detail the effects of the coupling coefficient between the atom and high-frequency mode photons and the three-wave mixing mediation coefficient on CPB and 2PB. The results of numerical calculations were compared with the analytical conditions, and they were consistent. The specific contributions of this study are as follows:
(i) Proposed a nonlinear system that can simultaneously realize CPB and 2PB, providing an alternative for realizing single-photon and two-photon sources.
(ii) Realized CPB in three resonant cavities with different frequencies simultaneously.
(iii) Achieved the transition between CPB and 2PB by adjusting the coupling coefficient between the atom and high-frequency mode photons under certain driving fields and photon frequencies.
(iiii) Confirmed that the introduction of the Jaynes-Cummings model can enhance the capability of the three-wave mixing system to realize CPB and 2PB.

\section{Physical Model and Analytical Analysis}
\label{sec:2}

In optical systems, the three-wave mixing process has been extensively studied theoretically and verified experimentally \cite{33,60}. In a recent study, a triply resonant mirroring resonator made of the nonlinear material aluminum nitrite $(AlN)$ successfully achieved three-wave mixing conversion \cite{61}. 
In this paper, we construct a three-wave mixing system embedded with a two-level atom in a high-frequency cavity. As shown in Fig. 1, the three-wave mixing process involves a high-frequency mode photon $a$, with frequency $\omega_{a}$, being converted, mediated by three-wave mixing, into two low-frequency mode photons $b$ and $c$, with frequencies $\omega_{b}$ and $\omega_{c}$, respectively.

\begin{center}
\includegraphics[scale=0.80]{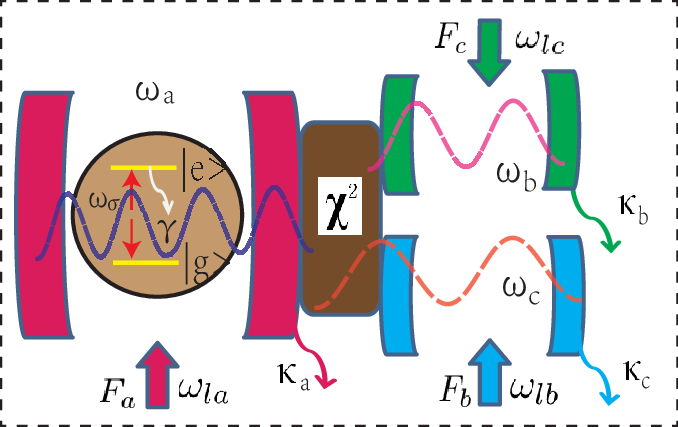}\\[5pt]  
\parbox[c]{13.0cm}{\footnotesize{\bf Fig.~1.}  A schematic diagram of a three-wave mixing system embedded with a two-level atomic system in a high-frequency cavity. Here, $\omega_a$, $\omega_b$, and $\omega_c$ represent the frequencies of photons in modes a, b, and c respectively, while $\omega_\sigma$ denotes the transition frequency between the ground state $|g\rangle$ and the excited state $|e\rangle$ of the atom. $F_a$, $F_b$, and $F_c$ denote the driving strengths of different mode photons, with driving frequencies $\omega_{la}$, $\omega_{lb}$, and $\omega_{lc}$ respectively. $\kappa_a$, $\kappa_b$, and $\kappa_c$ represent the decay rates of different cavities, while $\gamma$ denotes the decay rate of the atom.}
\label{Fig.1}
\end{center}

The specific process can be described by the following Hamiltonian\cite{33,60,61}:(setting $\hbar=1$)
\begin{eqnarray}
\hat{H_0}=\omega_a\hat{a}^{\dag}\hat{a}\
+\omega_\sigma{\sigma}^{\dag}{\sigma}+\omega_b\hat{b}^{\dag}\hat{b}+\omega_c\hat{c}^{\dag}\hat{c}+J(\hat{a}^{\dag}{\sigma}+{\sigma}^{\dag}\hat{a})
+g(\hat{a}^{\dag}\hat{b}\hat{c}+\hat{b}^{\dag}\hat{c}^{\dag}\hat{a}).
\label{01}
\end{eqnarray}
The symbols $\hat{a}(\hat{a}^{\dag})$, $\hat{b}(\hat{b}^{\dag})$ and $\hat{c}(\hat{c}^{\dag})$ represent the annihilation (creation) operators of modes $a$, $b$ and $c$, respectively, while $\sigma^{\dag}\sigma$ is the raising (lowering) operator of the two-level atom, and $J$ represents the nonlinear coupling coefficient between the atom and the high-frequency mode photon. And $g$ represents the coefficient governing the interactions in three-wave mixing. To maximize the effectiveness of these interactions, it's crucial to uphold the energy conservation principle, expressed as
$\omega_a=\omega_b+\omega_c=\omega_\sigma$. Considering the implementation of photon blockade involves converting a Poissonian light source into a sub-Poissonian one, external driving is necessary. Here, we opt for a three-driving mode, and the Hamiltonian is as follows:
\begin{eqnarray}
\hat{H} = \hat{H}_0 +  F_a (\hat{a}^\dag e^{-i\omega_{la} t} + \hat{a} e^{i\omega_{la} t})+  F_b (\hat{b}^\dag e^{-i\omega_{lb} t} + \hat{b} e^{i\omega_{lb} t})+ F_c (\hat{c}^\dag e^{-i\omega_{lc} t} + \hat{b} e^{i\omega_{lc} t}). \label{02}
\end{eqnarray}
Here,
$\omega_{la}$, $\omega_{lb}$ and $\omega_{lc}$ represent the driving frequencies of different mode photons, while the driving intensities are denoted by $F_a$, $F_b$ and $F_c$, respectively. When we only drive the high-frequency mode, the Hamiltonian exhibits a down-conversion process. When we drive the two low-frequency modes, it exhibits an up-conversion process. Under the three-driving mode, both frequency conversion modes coexist. Therefore, the current system can achieve CPB on all three photon modes.
For convenience, here we consider the rotating frame, and the effective Hamiltonian can be expressed as:
\begin{eqnarray}
\hat{H}_{d}&=&\Delta_a\hat{a}^{\dag}\hat{a}\
+\Delta_\sigma{\sigma}^{\dag}{\sigma}+\Delta_b\hat{b}^{\dag}\hat{b}+\Delta_c\hat{c}^{\dag}\hat{c}+J(\hat{a}^{\dag}{\sigma}+{\sigma}^{\dag}\hat{a})+g(\hat{a}^{\dag}\hat{b}\hat{c}+\hat{b}^{\dag}\hat{c}^{\dag}\hat{a})\nonumber\\
&&+F_a(\hat{a}^{\dag}+\hat{a})+F_b(\hat{b}^{\dag}+\hat{b})+F_c(\hat{c}^{\dag}+\hat{c}),
\label{03}
\end{eqnarray}
where  $\Delta_a=\omega_a-\omega_{la}$, $\Delta_b=\omega_b-\omega_{lb}$, $\Delta_c=\omega_c-\omega_{lc}$ and $\Delta_\sigma=\Delta_a$.
The Hamiltonian of the system can be represented in matrix form for different subspaces. Typically, achieving photon blockade requires the system to satisfy the condition of weak driving. Under this condition, interactions between subspaces can be neglected, allowing us to diagonalize the matrices corresponding to each subspace to obtain the eigenvalues.
Here, we denote the Fock state basis of the system as $|g(e),m,n,p\rangle$, where $m$, $n$, and $p$ represent the photon numbers in modes a, b, and c respectively, and $g(e)$ represents the ground (excited) state of the two-level atom. First, we choose $|g,1,0,0\rangle$, $|e,0,0,0\rangle$, and $|g,0,1,1\rangle$ form a closed subspace. We expand the Hamiltonian using these three bases, and the expanded Hamiltonian can be represented in the following matrix form:

\begin{eqnarray}
{H^{(1)}}=
\begin{bmatrix}
\Delta_a & J & g\\
 J & \Delta_\sigma & 0\\
 g & 0 & \Delta_b+\Delta_c
\end{bmatrix}.
 \label{04}
\end{eqnarray}

Diagonalizing the above matrix, we can obtain the two eigenfrequencies $\omega_{+}^{(1)}$ and $\omega_{-}^{(1)}$ of the first excited state of this coupled system, which can be represented as:
\begin{eqnarray}
\omega^{(1)}_{\pm}=\Delta_a\pm\sqrt{g^{2}+J^{2}}, & \omega^{(1)}_{0}=\Delta_a.
\label{05}
\end{eqnarray}
Here, by setting the eigenvalue of the first excited state of the system to zero, we obtain the conditions for achieving CPB in the system, as follows:
\begin{eqnarray}
\Delta_a = \Delta_b+\Delta_c = \pm\sqrt{g^{2}+J^{2}}.
\label{06}
\end{eqnarray}
The energy level transitions under the above conditions are shown in Fig. 2(a), where $\omega^{(1)}_{\pm}$ and $\omega^{(1)}_{0}$ are derived from Eq. (5) without considering the driving frequency.
Similarly, if we choose $|g,2,0,0\rangle$, $|e,1,0,0\rangle$, $|e,0,1,1\rangle$, $|g,0,2,2\rangle$, and $|g,1,1,1\rangle$ as the closed subspace, we can obtain the Hamiltonian of the system in this subspace. By diagonalizing it, we can obtain the eigenfrequencies of the second excited state of the system. The matrix form of the Hamiltonian in this subspace is as follows:

\begin{eqnarray}
{H^{(2)}}=
\begin{bmatrix}
2\Delta_a & \sqrt{2}J & 0 & 0 & \sqrt{2}g\\
\sqrt{2}J & \Delta_a+\Delta_{\sigma} & g & 0 & 0\\
0 & g & \Delta_{\sigma}+\Delta_b+\Delta_c & 0 & J\\
0 & 0 & 0 & 2(\Delta_b+\Delta_c) & 2g\\
\sqrt{2}g & 0 & J & 2g & \Delta_a+\Delta_b+\Delta_c
\end{bmatrix}.
 \label{07}
\end{eqnarray}

The result after diagonalization is:
\begin{eqnarray}
\omega^{(2)}_{\pm\pm}=2\Delta_a\pm\frac{\sqrt{2}}{2}\sqrt{A\pm\sqrt{B}}, & \omega^{(2)}_{0}=2\Delta_a,
\label{08}
\end{eqnarray}
where $A=7g^2+3J^2$ and $B=\sqrt{25g^4+26g^2J^2+J^4}$. By setting the eigenvalue $\omega^{(2)}_{\pm\pm}$ of the second excited state to zero, we obtain the analytical conditions for achieving two-photon blockade in the system, which are as follows:
\begin{eqnarray}
\Delta_{a1} = \Delta_b+\Delta_c = \pm\frac{\sqrt{2}}{4}\sqrt{A+\sqrt{B}},\nonumber\\
\Delta_{a2} = \Delta_b+\Delta_c = \pm\frac{\sqrt{2}}{4}\sqrt{A-\sqrt{B}}.
\label{09}
\end{eqnarray}
The energy-level transition process under the conditions specified in Eq. (9) is shown in Fig. 2(b), where
$\omega^{(2)}_{\pm\pm}$ and $\omega^{(2)}_{0}$ represent the energy levels of the two-photon excited system for mode $a$.
When any of the analytical conditions in Eq. (6) are met, the system will occupy the first excited state, leading to the occurrence of CPB. Similarly, when any of the analytical conditions in Eq. (9) are satisfied, the system will occupy the second excited state, i.e., the transition $0 \rightarrow \omega_{\pm}^{(2)}$ is resonant, and the transition $0 \rightarrow \omega_{\pm}^{(1)}$ is detuned, resulting in the occurrence of 2PB. Next, we will study CPB and 2PB in the system through numerical simulations.

\begin{center}
\includegraphics[scale=0.70]{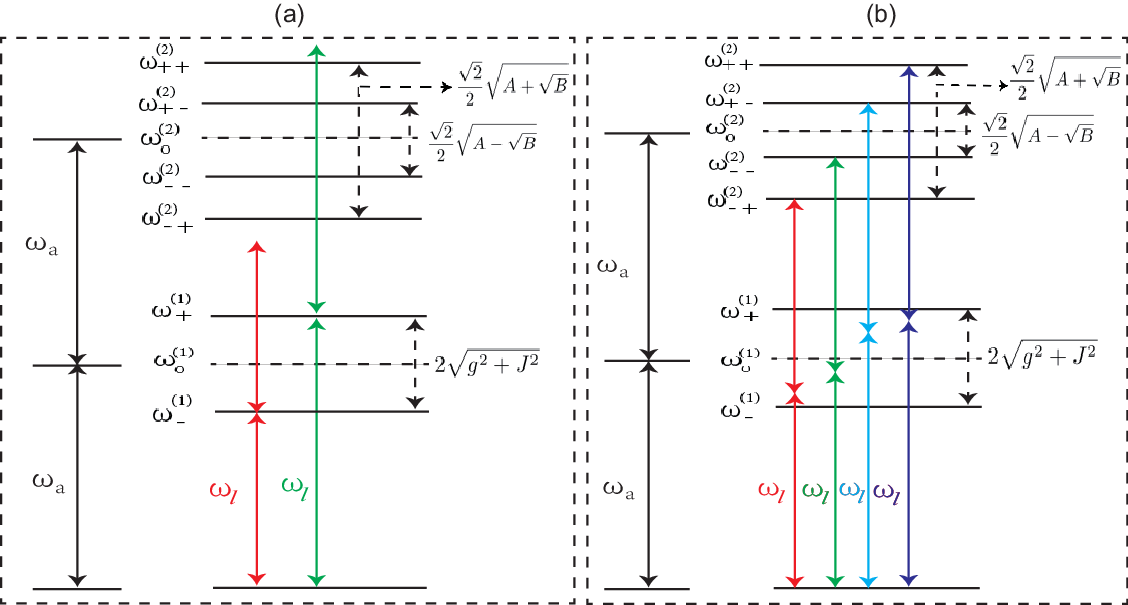}\\[5pt]  
\parbox[c]{15.0cm}{\footnotesize{\bf Fig.~2.}  (a) The energy level schematic diagram illustrating the realization of CPB in mode a, under the condition where the transition $0\rightarrow\omega_{\pm}^{(1)}$ is resonant, and the transition $\omega_{\pm}^{(1)}\rightarrow\omega_{\pm}^{(2)}$ is detuned.(b) The energy level transition diagram illustrating the realization of two-photon blockade (2PB) in mode a, under the condition where the transition $0\rightarrow\omega_{\pm}^{(2)}$ is resonant, and the transition $0\rightarrow\omega_{\pm}^{(1)}$ is detuned.}
\label{Fig.2}
\end{center}

\section{Numerical Analysis of Photon Blockade}
\label{sec:3}
The dynamics of this open quantum system is governed by the master equation, i.e.,
\begin{eqnarray}
\frac{\partial\hat{\rho}}{\partial t}&=&-i[\hat{H},\rho]+\frac{\kappa_a}{2}(\bar{n}_{ta}+1)(2\hat{a}\hat{\rho}\hat{a}^\dag+\frac{1}{2}\hat{a}^\dag\hat{a}\hat{\rho}+\frac{1}{2}\hat{\rho}\hat{a}^\dag\hat{a})\nonumber\\
&&+\frac{\kappa_b}{2}(\bar{n}_{tb}+1)(2\hat{b}\hat{\rho}\hat{b}^\dag+\frac{1}{2}\hat{b}^\dag\hat{b}\hat{\rho}+\frac{1}{2}{\rho}\hat{b}^\dag{b})\nonumber\\
&&+\frac{\kappa_c}{2}(\bar{n}_{tc}+1)(2\hat{c}\hat{\rho}\hat{c}^\dag+\frac{1}{2}\hat{c}^\dag\hat{c}\hat{\rho}+\frac{1}{2}{\rho}\hat{c}^\dag{c})\nonumber\\
&&+\frac{\gamma}{2}(\bar{n}_{t\sigma}+1)(2{\sigma}{\rho}{\sigma}^\dag+\frac{1}{2}{\sigma}^\dag{\sigma}\hat{\rho}+\frac{1}{2}{\rho}{c}^\dag{c})\nonumber\\
&&+\frac{\kappa_a}{2}\bar{n}_{ta}(2\hat{a}^\dag\hat{\rho}\hat{a}+\frac{1}{2}\hat{a}\hat{a}^\dag\hat{\rho}+\frac{1}{2}\hat{\rho}\hat{a}\hat{a}^\dag)\nonumber\\
&&+\frac{\kappa_b}{2}\bar{n}_{tb}(2\hat{b}^\dag\hat{\rho}\hat{b}+\frac{1}{2}\hat{b}\hat{b}^\dag\hat{\rho}+\frac{1}{2}\hat{\rho}\hat{b}\hat{b}^\dag) \nonumber\\
&&+\frac{\kappa_c}{2}\bar{n}_{tc}(2\hat{c}^\dag\hat{\rho}\hat{c}+\frac{1}{2}\hat{c}\hat{c}^\dag\hat{\rho}+\frac{1}{2}\hat{\rho}\hat{c}\hat{c}^\dag) \nonumber\\
&&+\frac{\gamma}{2}\bar{n}_{t\sigma}(2{\sigma}{\rho}{\sigma}^\dag+\frac{1}{2}{\sigma}^\dag{\sigma}\hat{\rho}+\frac{1}{2}\hat{\rho}{\sigma}^\dag{\sigma}).\nonumber\\
\label{10}
\end{eqnarray}
Where $\kappa_a$ represents the decay rate of the high-frequency mode photon a, while $\kappa_b$ and $\kappa_c$ respectively represent the decay rates of the two low-frequency mode photons b and c. $\gamma$ denotes the spontaneous emission rate of the two-level atom. $\bar{n}_{th}$ represents the average number of thermal photons, expressed as  $\bar{n}_{ti}=\{\exp{[\hbar\omega_i/(\kappa_BT_i)-1]}\}^{-1}$, where $\kappa_B$ is the Boltzmann constant, and $T$ is the thermodynamic temperature of the reservoir in thermal equilibrium. The statistical properties of photons are described by the zero-delay-time correlation function $g_{i}^{(n)}(0)$ in the steady state. Therefore, we need the steady-state density operator $\hat{\rho}_s$, which can be obtained by setting $\frac{\partial\hat{\rho}}{\partial t}=0$. In the current system, we will discuss the CPB of three photon modes a, b, and c, as well as the 2PB of the high-frequency mode. The zero-delay-time correlation function is defined as follows:
\begin{eqnarray}
g_i^{(n)}(0)=\frac{\langle
\hat{i}^{\dag n}\hat{i}^n\rangle}
{\langle \hat{i}^{\dag}\hat{i}\rangle^n}.
\label{11}
\end{eqnarray}
Here, $i=a$, $b$, or $c$, and $n=2$ or $3$. When $g_i^{(2)}(0) < 1$, it implies that photons in mode $i$ exhibit sub-Poissonian statistics, i.e., CPB occurs in mode $i$. If $g_a^{(2)}(0) \geq 1$ and $g_a^{(3)}(0) < 1$ hold simultaneously, it indicates that photons in mode $a$ undergo 2PB. The brightness at the occurrence of photon blockade is represented by the average photon number $N_{i} = \langle \hat{i}^{\dag}\hat{i} \rangle = \text{Tr}(\hat{\rho}_s\hat{i}^{\dag}\hat{i})$.

\subsection{Numerical Results of the system in implementing CPB}
\begin{center}
\includegraphics[scale=0.60]{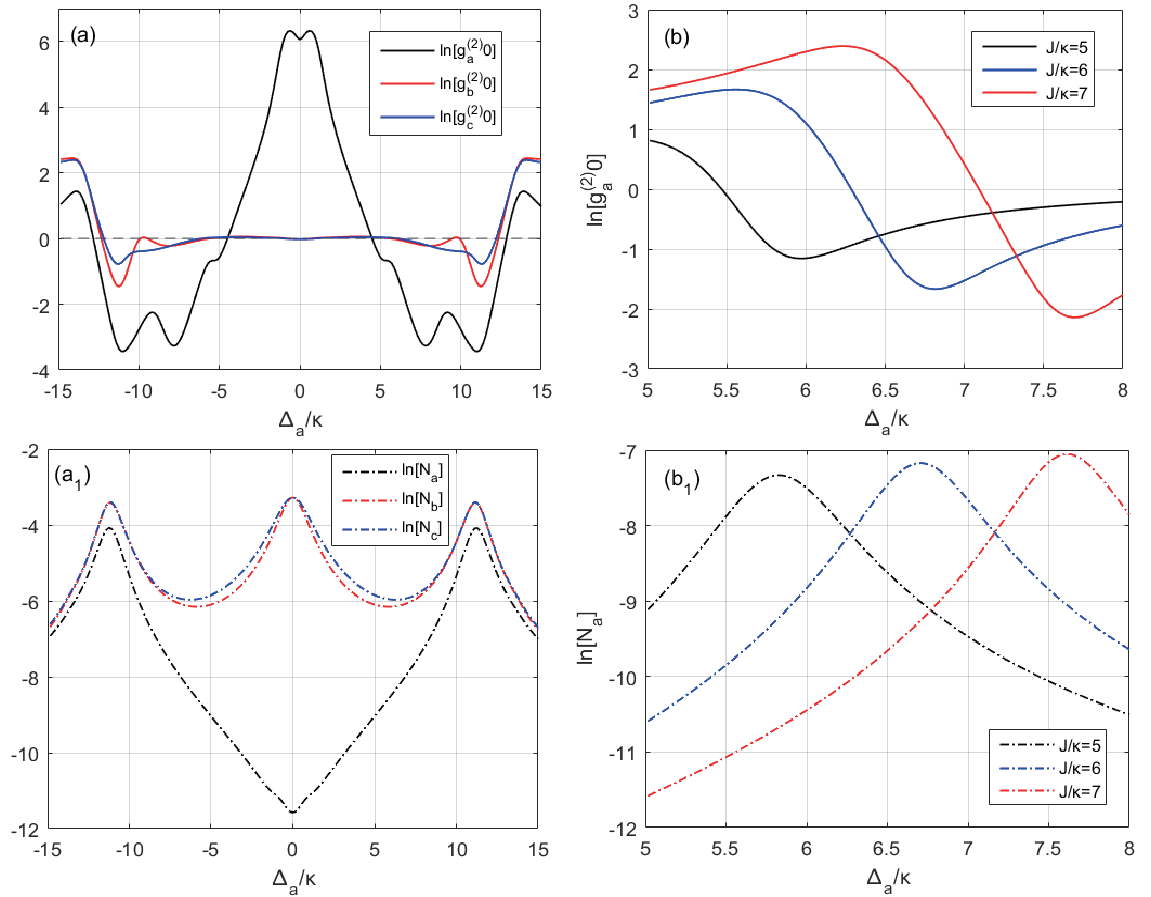}\\[5pt]  
\parbox[c]{15.0cm}{\footnotesize{\bf Fig.~3.}  (a)Logarithmic plot of the $g_{a}^{(2)}(0)$, $g_{b}^{(2)}(0)$, and $g_{c}^{(2)}(0)$ as functions of $\Delta_{a}/\kappa$ for $\Delta_{b}/\kappa=\frac{2}{3}\Delta_{a}/\kappa$, $\Delta_{c}/\kappa=\frac{1}{3}\Delta_{a}/\kappa$, $\Delta_{\sigma}/\kappa=\Delta_{a}/\kappa$ $g/\kappa=11$, $J/\kappa=2$, $F_{a}/\kappa=0.1$, $F_{b}/\kappa=0.05$ and $F_{c}/\kappa=0.05$.  $(a_{1})$ The logarithmic values of $N_{a}$, $N_{b}$, and $N_{c}$ under the same parameters and coordinates as in Figure (a). (b)Curves of the logarithmic values of $g_{a}^{(2)}(0)$ as a function of $\Delta_{a}/\kappa$ were plotted for different values of $J/\kappa$, with other parameters set as $\Delta_{b}/\kappa=\frac{2}{3}\Delta_{a}/\kappa$, $\Delta_{c}/\kappa=\frac{1}{3}\Delta_{a}/\kappa$, $\Delta_{\sigma}/\kappa=\Delta_{a}/\kappa$, $g/\kappa=3$, $F_{a}/\kappa=0.02$, $F_{b}/\kappa=0.01$ and $F_{c}/\kappa=0.01$. $(b_{1})$ The logarithmic values of $N_{a}$ under the same parameters and coordinates as in Figure (b).}
\label{Fig.3}
\end{center}

We first plotted the relationship curves of $g_{n}^{(2)}(0)$ and $N_{n}$ under different system parameters as numerical results to discuss CPB in different photon modes($n=a,b,c$). The values of $g_{n}^{(2)}(0)$  and $N_{n}$ were obtained by substituting the effective Hamiltonian $H_d$ shown in Equation (3) into the master equation shown in Equation (10). The Hilbert space of the system for the three photon modes was truncated to five dimensions, and to two dimensions for the atom. Figure. 1 presents the numerical results of the CPB effect, assuming a zero-temperature environment $\bar{n}_{th}=0$ and equal decay rates for the cavity modes, i.e., $\kappa_a = \kappa_b = \kappa_c = \kappa$. All system parameters are rescaled with respect to the decay rate $\kappa$.

In Figure. 3(a), we present the relationship curves of $g_{a}^{(2)}(0)$, $g_{b}^{(2)}(0)$, and $g_{c}^{(2)}(0)$ with respect to $\Delta_{a}/\kappa$. The other parameters are set as $\Delta_{b}/\kappa = \frac{2}{3}\Delta_{a}/\kappa$, $\Delta_{c}/\kappa = \frac{1}{3}\Delta_{a}/\kappa$, $\Delta_{\sigma}/\kappa = \Delta_{a}/\kappa$, $g/\kappa = 11$, $J/\kappa = 2$, $F_{a}/\kappa = 0.1$, $F_{b}/\kappa = 0.05$, and $F_{c}/\kappa = 0.05$. The results indicate that CPB effects occur in cavity modes a, b, and c, and the optimal blockade points align with the analytical conditions shown in Equation (6). Comparatively, the CPB effect has the highest purity in the high-frequency cavity. This may be due to the nature of three-wave mixing mediated by the three driving modes, which allows for bidirectional parametric conversion processes. However, the conversion process from one high-frequency mode photon to two low-frequency mode photons of different frequencies is easier to achieve, whereas converting two low-frequency mode photons of different frequencies into a specific high-frequency mode photon is more complex. Additionally, according to the analytical conditions shown in Equation (6) ($\Delta_a = \Delta_b + \Delta_c = \pm\sqrt{g^{2} + J^{2}}$), the energy level splitting of the high-frequency mode is the largest, making it most significantly affected by nonlinear effects. To achieve a high-quality single-photon source, the system must not only realize high-purity PB effects, indicated by a smaller value of $g^{(2)}(0)$, but also possess a sufficiently high brightness value. Brightness is defined as the average photon number  $N_{i} = \langle \hat{i}^{\dag}\hat{i} \rangle = \text{Tr}(\hat{\rho}_s\hat{i}^{\dag}\hat{i})$, and $i=a/b/c$, which can be obtained by numerically solving the master equation. In Figure. 3$(a_{1})$, we present the logarithmic values of $N_{a}$,  $N_{b}$, and $N_{c}$ as functions of $\Delta_{a}/\kappa$, with other parameters set identically to those in Figure. 3(a). The results show that all three photon modes exhibit larger average photon numbers, indicating higher brightness, within the optimal blockade regions.

Next, we will further discuss the CPB in the high-frequency cavity and the impact of introducing a two-level atom, which couples with the high-frequency mode photon via the coupling coefficient $J/\kappa$. In Fig. 3(b), we plot the logarithmic values of $g_{a}^{(2)}(0)$ as a function of $\Delta_{a}/\kappa$ for different values of $J/\kappa$, with parameters set as $\Delta_{b}/\kappa=\frac{2}{3}\Delta_{a}/\kappa$, $\Delta_{c}/\kappa=\frac{1}{3}\Delta_{a}/\kappa$, $\Delta_{\sigma}/\kappa=\Delta_{a}/\kappa$, $g/\kappa=3$, $F_{a}/\kappa=0.02$, $F_{b}/\kappa=0.01$ and $F_{c}/\kappa=0.01$. The results indicate that with the three-wave mixing coefficient $g/\kappa$ fixed, as $J/\kappa$ increases, the optimal blockade points shift towards higher values of $\Delta_{a}/\kappa$, consistent with the analytical solution shown in Eq. (6).  Moreover, as $J/\kappa$ increases, the minimum value of $g_{a}^{(2)}(0)$ decreases, indicating an increase in the purity of single photons in the high-frequency cavity. This suggests that the introduction of a two-level atom can enhance the capability of the three-wave mixing system to achieve CPB. In Fig. 3$(b_{1})$, we present the logarithmic values of $N_{a}$ as a function of
$\Delta_{a}/\kappa$, with the same parameter settings as in Fig. 3(b). The results show that the optimal blockade regions exhibit higher numbers of high-frequency mode photons, which is beneficial for the realization of a single-photon source.

\subsection{Numerical Results of the system in implementing 2PB}

Next, we will perform numerical calculations of two-photon blockade (2PB) in the high-frequency cavity and discuss the impact of the coupling coefficient between the two-level atom and the high-frequency mode photon on 2PB. According to the analytical results, when any of the analytical conditions shown in Eq. (9) are satisfied, the system will occupy the second excited state, meaning that the transition the $0 \rightarrow \omega_{\pm}^{(2)}$ is resonant,  while the transition $0 \rightarrow \omega_{\pm}^{(1)}$ is detuned. In Figs. 4(a), (b), and (c), we plot the logarithmic values of $g_{a}^{(2)}(0)$ and $g_{a}^{(3)}(0)$ as functions of $\Delta_{a}/\kappa$ for different values of $J/\kappa$, where $g/\kappa=8$, $\Delta_{b}/\kappa=\frac{2}{3}\Delta_{a}/\kappa$, $\Delta_{c}/\kappa=\frac{1}{3}\Delta_{a}/\kappa$, $\Delta_{\sigma}/\kappa=\Delta_{a}/\kappa$, $F_{a}/\kappa=0.1$, and $F_{b}/\kappa=F_{c}/\kappa=0$. According to the results in Fig. 4(a), when $J/\kappa=2$, the region where $g_a^{(2)}(0) \geq 1$ and $g_a^{(3)}(0) < 1$ occurs around $9.508<\Delta_{a}/\kappa<10.5$. Substituting the current parameters into the analytical condition for $\Delta_{a1} = \Delta_b+\Delta_c = \pm\frac{\sqrt{2}}{4}\sqrt{A+\sqrt{B}}$ shown in Eq. (9), the second excited state should occur at $\Delta_{a}/\kappa=9.94$, which is consistent with the numerical results. Therefore, the current system can exhibit 2PB in the high-frequency cavity, and the analytical solution matches the numerical solution. In Fig. 4(b) and (c), where $J/\kappa$ is set to 4 and 6 respectively, 2PB occurs around $9.991<\Delta_{a}/\kappa<10.91$ and $10.77<\Delta_{a}/\kappa<11.57$. According to the analytical conditions, the optimal blockade regions should occur at $\Delta_{a}/\kappa=10.3$ and $\Delta_{a}/\kappa=10.94$. Similar to the results in Fig. 4(a), the system can achieve 2PB in the high-frequency mode, and the numerical solutions align with the analytical solutions. By comparing the results of Fig. 4(a), (b), and (c), we observe that with a fixed three-wave mixing coefficient $g/\kappa$, as the coupling coefficient $J/\kappa$ between the high-frequency mode photon and the two-level atom increases, the value of $g_{a}^{(3)}(0)$ at the optimal blockade point gradually decreases. This indicates that the purity of two-photon blockade in the blockade region gradually increases. This suggests that the introduction of a two-level atom is beneficial for the occurrence of 2PB in the system.

Next, we will discuss the relationship between the parameters $J/\kappa$ and $g/\kappa$ when the system achieves 2PB in the high-frequency mode. In Fig. 4(d), (e), and (f), we plot the logarithmic values of $g_{a}^{(2)}(0)$ and $g_{a}^{(3)}(0)$ as functions of $g/\kappa$ for different values of $J/\kappa$. Other parameters are set to $\Delta_{a}/\kappa=7$, $\Delta_{b}/\kappa=\frac{2}{3}\Delta_{a}/\kappa$, $\Delta_{c}/\kappa=\frac{1}{3}\Delta_{a}/\kappa$, $\Delta_{\sigma}/\kappa=\Delta_{a}/\kappa$, $F_{a}/\kappa=0.1$, and $F_{b}/\kappa=F_{c}/\kappa=0$. When $J/\kappa$ is 1, 2, and 3, 2PB occurs in the regions of  $5.2<g/\kappa<5.81$,  $5.08<g/\kappa<5.68$ and  $4.84<g/\kappa<5.4$, respectively. Substituting the current parameters into the analytical condition $\Delta_{a2} = \Delta_b+\Delta_c = \pm\frac{\sqrt{2}}{4}\sqrt{A-\sqrt{B}}$ from Eq. (9), we find that the second excited state should occur at $g/\kappa=5.66$, $g/\kappa=5.55$ and $g/\kappa=5.34$. The numerical results show that as $J/\kappa$ increases, the value of $g_{a}^{(3)}(0)$ at the optimal blockade point gradually decreases. Consistent with the conclusions from Fig. 4(a), (b), and (c), the analytical solution aligns with the numerical solution, and the introduction of the two-level atom enhances the system's capability to achieve 2PB in the high-frequency cavity.

\begin{center}
\includegraphics[scale=0.70]{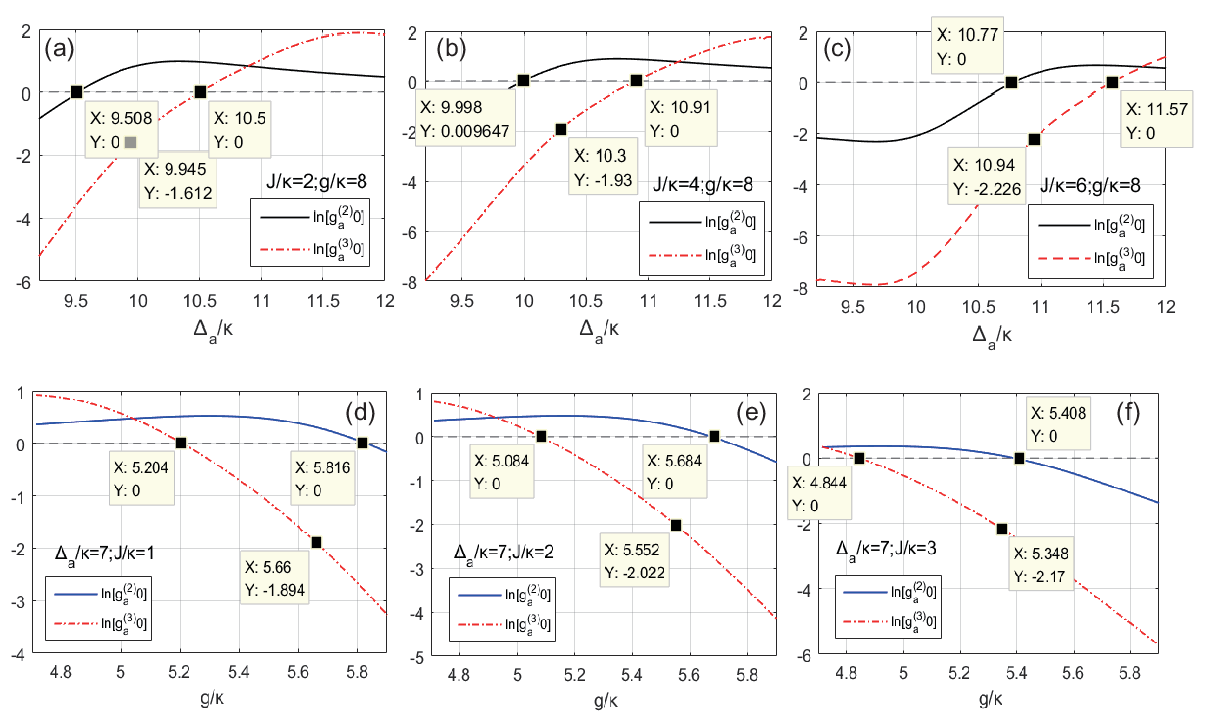}\\[5pt]  
\parbox[c]{13.0cm}{\footnotesize{\bf Fig.~4.} (a), (b), (c) Under different values of  $J/\kappa$, plot the logarithmic curves of $g_{a}^{(2)}(0)$ and $g_{a}^{(3)}(0)$ versus $\Delta_{a}/\kappa$, where $g/\kappa=8$, $\Delta_{b}/\kappa=\frac{2}{3}\Delta_{a}/\kappa$, $\Delta_{c}/\kappa=\frac{1}{3}\Delta_{a}/\kappa$, $\Delta_{\sigma}/\kappa=\Delta_{a}/\kappa$, $F_{a}/\kappa=0.1$, and $F_{b}/\kappa=F_{c}/\kappa=0$. (d), (e), (f) Under different values of  $J/\kappa$, plot the logarithmic curves of $g_{a}^{(2)}(0)$ and $g_{a}^{(3)}(0)$ versus $g/\kappa$, where $\Delta_{a}/\kappa=7$, $\Delta_{b}/\kappa=\frac{2}{3}\Delta_{a}/\kappa$, $\Delta_{c}/\kappa=\frac{1}{3}\Delta_{a}/\kappa$, $\Delta_{\sigma}/\kappa=\Delta_{a}/\kappa$, $F_{a}/\kappa=0.1$, and $F_{b}/\kappa=F_{c}/\kappa=0$.}
\label{Fig.4}
\end{center}
The process of realizing the photon blockade effect involves converting a driven light source with Poissonian statistics into a source with sub-Poissonian statistics. Therefore, the presence of a driving field is a necessary condition, and the photon blockade effect generally occurs under weak driving conditions. To discuss the impact of the driving coefficient of high-frequency mode photons on 2PB, we plot the logarithmic values of $g_{a}^{(2)}(0)$ and $g_{a}^{(3)}(0)$ as functions of $F_a/\kappa$ in Fig. 5(a). Other parameters are set as $\Delta_{a}/\kappa=8$, $g/\kappa=6.2$, $J/\kappa=3$, $\Delta_{b}/\kappa=6$, $\Delta_{c}/\kappa=2$, $\Delta_{\sigma}/\kappa=8$, and $F_{b}/\kappa=F_{c}/\kappa=0$. The values of $g/\kappa$, $J/\kappa$, and $\Delta_{a}/\kappa$ are set to satisfy the 2PB analytical condition ($\Delta_{a1} = \Delta_b+\Delta_c = \pm\frac{\sqrt{2}}{4}\sqrt{A+\sqrt{B}}$). According to the calculation results, under the current parameters, 2PB can occur in the region where $F_a/\kappa<0.4$.

In Fig. 5(b), we plot the logarithmic values of $g_{a}^{(2)}(0)$ and $g_{a}^{(3)}(0)$ as functions of $J/\kappa$. Other parameters are set as  $\Delta_{a}/\kappa=8$, $g/\kappa=6.2$, $\Delta_{b}/\kappa=6$, $\Delta_{c}/\kappa=2$, $\Delta_{\sigma}/\kappa=8$, $F_{a}/\kappa=0.1$, and $F_{b}/\kappa=F_{c}/\kappa=0$. According to the calculation results, 2PB can occur within the range of $J/\kappa=\pm3$, which is consistent with the analytical condition for 2PB. Additionally, by varying the values of $J/\kappa$, the PB effect in the high-frequency cavity can be switched between 2PB and CPB. This means that with the current system, when realizing a sub-Poissonian light source, we can achieve the conversion between a single-photon source and a two-photon source by adjusting the coupling coefficient between the atom and the high-frequency mode photon.
\begin{center}
\includegraphics[scale=0.70]{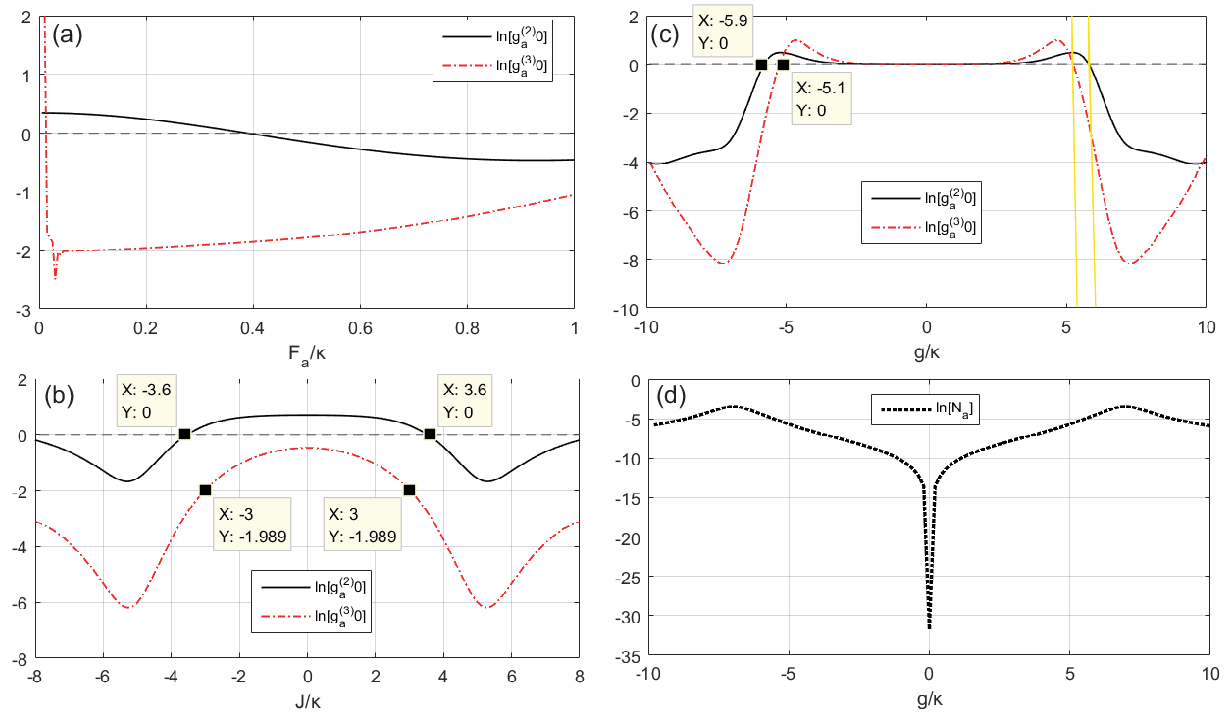}\\[5pt]  
\parbox[c]{15.0cm}{\footnotesize{\bf Fig.~5.} (a)Logarithmic plot of the $g_{a}^{(2)}(0)$ and $g_{a}^{(3)}(0)$  as functions of $F_{a}/\kappa$ for $\Delta_{a}/\kappa=8$, $g/\kappa=6.2$, $J/\kappa=3$, $\Delta_{b}/\kappa=6$, $\Delta_{c}/\kappa=2$, $\Delta_{\sigma}/\kappa=8$, and $F_{b}/\kappa=F_{c}/\kappa=0$. (b) Logarithmic plot of the $g_{2}^{(2)}(0)$ and $g_{a}^{(3)}(0)$  as functions of $J/\kappa$ for $\Delta_{a}/\kappa=8$, $g/\kappa=6.2$, $\Delta_{b}/\kappa=6$, $\Delta_{c}/\kappa=2$, $\Delta_{\sigma}/\kappa=8$, $F_{a}/\kappa=0.1$, and $F_{b}/\kappa=F_{c}/\kappa=0$. (c) Logarithmic plot of the $g_{a}^{(2)}(0)$ and $g_{a}^{(3)}(0)$  as functions of $g/\kappa$ for $\Delta_{a}/\kappa=7$, $J/\kappa=1$, $\Delta_{b}/\kappa=3$, $\Delta_{c}/\kappa=4$, $\Delta_{\sigma}/\kappa=7$, $F_{a}/\kappa=0.05$, and $F_{b}/\kappa=F_{c}/\kappa=0$.(d) Logarithmic plot of the $N_{a}$  as functions of $g/\kappa$ for $\Delta_{a}/\kappa=7$, $J/\kappa=1$, $\Delta_{b}/\kappa=3$, $\Delta_{c}/\kappa=4$, $\Delta_{\sigma}/\kappa=7$, $F_{a}/\kappa=0.05$, and $F_{b}/\kappa=F_{c}/\kappa=0$.}
\label{Fig.5}
\end{center}
To further discuss the impact of the three-wave mixing coefficient $g/\kappa$ on 2PB, in Fig. 5(c), we plot the logarithmic values of $g_{a}^{(2)}(0)$ and $g_{a}^{(3)}(0)$ as functions of $g/\kappa$. Other parameters are set to $\Delta_{a}/\kappa=7$, $J/\kappa=1$, $\Delta_{b}/\kappa=3$, $\Delta_{c}/\kappa=4$, $\Delta_{\sigma}/\kappa=7$, $F_{a}/\kappa=0.05$, and $F_{b}/\kappa=F_{c}/\kappa=0$. 2PB occurs in the regions of $-5.9<g/\kappa<-5.1$ and $5.1<g/\kappa<5.9$, which is consistent with the analytical condition $\Delta_{a2} = \Delta_b+\Delta_c = \pm\frac{\sqrt{2}}{4}\sqrt{A-\sqrt{B}}$ from Eq. (9). Moreover, similar to the effect of the parameter $J$ in Fig. 5(b), varying the coupling coefficient $g/\kappa$ can achieve the transition between 2PB and CPB in the high-frequency cavity. Fig. 5(d) shows the logarithmic values of $N_a$ as a function of $g/\kappa$ with the same parameter settings as in Fig. 5(c). The numerical results indicate that the regions where CPB and 2PB occur have higher brightness values, which is beneficial for realizing sub-Poissonian light sources.

All the numerical calculations mentioned above demonstrate that the composite three-wave mixing system discussed can achieve both CPB and 2PB simultaneously. Moreover, there are numerous potential experimental schemes for realizing three-wave mixing induction as discussed in this model. For example, embedding a three-mode resonator with high second-order nonlinear materials \cite{62,63,64}, or employing circuit quantum electrodynamics schemes, among others. Furthermore, the parameter settings we adopted in the numerical calculations are realistic estimations. For instance, the coupling coefficient $J/\kappa=3$ between the two-level atom and the high-frequency cavity \cite{65}. The three-wave mixing coefficient $g/\kappa=10$ \cite{66}, and its value can be adjusted by tuning the zero-delay-time second-order or third-order correlation functions.

\section{Conclusion}
In conclusion, based on the aforementioned research findings, the following conclusions can be drawn: The investigated three-wave mixing system, with a high-frequency cavity embedding a two-level atom, can achieve both CPB and 2PB simultaneously. By analytically analyzing the system Hamiltonian, we obtained the analytical conditions for CPB and 2PB, which were subsequently validated through numerical calculations. The system not only achieves CPB across all three photon modes but also realizes 2PB within the high-frequency cavity. The incorporation of a two-level atom in the high-frequency cavity not only enhances the purity of CPB and 2PB realization but also allows for the transition between CPB and 2PB within the high-frequency mode by adjusting the coupling coefficient between the atom and high-frequency mode photons under specified driving intensity and detuning conditions. Discussions were also conducted regarding the range of driving field intensity required to achieve 2PB in the system. Furthermore, it was determined that the blockade region exhibits higher brightness during CPB and 2PB realization.

\section*{Acknowledgments}
This was was supported in part by the Jilin Provincial Natural Science Foundation, grant number 20240101305JC. JiLin Education Department Fund under Grant No. (Grant No.JJKH20230015KJ). Innovative Research Team of Baicheng Normal University, China (Grant No.IRTBCNU).

\bibliographystyle{unsrt}  
\bibliography{lin}

\end{document}